\documentclass[apj]{emulateapj}

\usepackage{amsmath}

\usepackage{epsf}

\renewcommand\email\texttt

\def\change#1{{#1}}

\newcommand\coords{{14:00:06\, +14:30:00 $\pm 15\arcsec$}} 
\def\spose#1{\hbox to 0pt{#1\hss}}
\def\lta{\mathrel{\spose{\lower 3pt\hbox{$\sim$}}
    \raise 2.0pt\hbox{$<$}}}
\def\gta{\mathrel{\spose{\lower 3pt\hbox{$\sim$}}
    \raise 2.0pt\hbox{$>$}}}

\begin{document} 

\slugcomment{\sc submitted to \it Astrophysical Journal Letters}
\shorttitle{\sc New Satellite in Bo{\"o}tes} 
\shortauthors{Belokurov et al.}

\title{A Faint New Milky Way Satellite in Bo{\"o}tes}

\author{V. Belokurov\altaffilmark{1},
D.\ B. Zucker\altaffilmark{1}, 
N.\ W. Evans\altaffilmark{1}, 
M.\ I. Wilkinson\altaffilmark{1},
M.\ J. Irwin\altaffilmark{1},
S.\ Hodgkin\altaffilmark{1},
D.\ M. Bramich\altaffilmark{1}, 
J.\ M. Irwin\altaffilmark{1},
G. Gilmore\altaffilmark{1}, 
B. Willman\altaffilmark{2},
S. Vidrih\altaffilmark{1},
H.\ J.\ Newberg\altaffilmark{3},
R.\ F.\ G.\ Wyse\altaffilmark{4},
M. Fellhauer\altaffilmark{1},
P.\ C. Hewett\altaffilmark{1}, 
N. Cole\altaffilmark{3},
E.\ F.\ Bell\altaffilmark{5},
T.\ C. Beers\altaffilmark{6},
C.\ M.\ Rockosi\altaffilmark{7},
B. Yanny\altaffilmark{8},
E.\ K.\ Grebel\altaffilmark{9},
D.\ P.\ Schneider\altaffilmark{10},
R.\ Lupton\altaffilmark{11},
J.\ C.\ Barentine\altaffilmark{12},
H.\ Brewington\altaffilmark{12},
J.\ Brinkmann\altaffilmark{12},
M.\ Harvanek\altaffilmark{12},
S.\ J.\ Kleinman\altaffilmark{12},
J.\ Krzesinski\altaffilmark{12,13},
D.\ Long\altaffilmark{12},
A.\ Nitta\altaffilmark{12},
J.\ A.\ Smith\altaffilmark{14}
S.\ A.\ Snedden\altaffilmark{12}
}

\altaffiltext{1}{Institute of Astronomy, University of Cambridge,
Madingley Road, Cambridge CB3 0HA, UK;\email{vasily,zucker,nwe@ast.cam.ac.uk}}
\altaffiltext{2}{Center for Cosmology and Particle Physics, Department of Physics, New York University, 4 Washington Place, New York, NY 10003}
\altaffiltext{3}{Rensselaer Polytechnic Institute, Troy, NY 12180}
\altaffiltext{4}{The Johns Hopkins University, 3701 San Martin Drive,
Baltimore, MD 21218}
\altaffiltext{5}{Max Planck Institute for Astronomy, K\"{o}nigstuhl
17, 69117 Heidelberg, Germany}
\altaffiltext{6}{Department of Physics and Astronomy, CSCE: Center for
the Study of Cosmic Evolution, and JINA: Joint Institute for Nuclear
Astrophysics, Michigan State University, East Lansing, MI 48824}
\altaffiltext{7}{Lick Observatory, University of California, Santa Cruz, CA 95064}
\altaffiltext{8}{Fermi National Accelerator Laboratory, P.O. Box 500,
Batavia, IL 60510}
\altaffiltext{9}{Astronomical Institute of the University of Basel,
Department of Physics and Astronomy, Venusstrasse 7,CH-4102 Binningen, 
Switzerland}
\altaffiltext{10}{Department of Astronomy and Astrophysics,
Pennsylvania State University, 525 Davey Laboratory, University Park,
PA 16802}
\altaffiltext{11}{Princeton University Observatory, Peyton Hall, Princeton, NJ 08544}
\altaffiltext{12}{Apache Point Observatory, P.O. Box 59, Sunspot, NM 88349}
\altaffiltext{13}{Mt. Suhora Observatory, Cracow Pedagogical University, ul. Podchorazych 2, 30-084 Cracow, Poland}
\altaffiltext{14 }{Los Alamos National Laboratory, ISR-4, MS D448, Los 
Alamos, NM 87545}


\begin{abstract}
In this Letter, we announce the discovery of a new satellite of the
Milky Way in the constellation of Bo\"otes at a distance of $\sim 60$
kpc. It was found in a systematic search for stellar overdensities in
the North Galactic Cap using Sloan Digital Sky Survey Data Release 5
(SDSS DR5).  The color-magnitude diagram shows a well-defined
turn-off, red giant branch, and extended horizontal branch.  Its
absolute magnitude is $M_V \sim -5\fm8$, which \change{makes it one of
the faintest galaxies known}.  The half-light radius is $\sim 220$
pc. The isodensity contours are elongated and have an irregular shape,
suggesting that Boo may be a disrupted dwarf spheroidal galaxy.
\end{abstract}


\section{Introduction}

The last few years have seen a number of discoveries of new satellite
companions to the Milky Way. \citet{Wi05} \change{systematically}
surveyed $\sim 5800$ square degrees of the Sloan Digital Sky Survey
(SDSS; York et al. 2000) and identified two strong candidates. The
first, now called Willman 1, is an unusually extended object with
properties intermediate between those of globular clusters and dwarf
galaxies~\citep{Wi05a}. The second proved to be a new dwarf spheroidal
(dSph) companion to the Milky Way, located in the constellation of
Ursa Major~\citep{Wi05b,Ke05}.

Very recently, \cite{Zu06} serendipitously discovered a stellar
overdensity in the ``Field of Streams''~\citep{Be06} -- a plot of the
halo substructure in the Galactic northern hemisphere derived from
SDSS Data Release 5 (DR5; Adelman-McCarthy et al. 2006). Closer
analysis revealed that this was a new dwarf spheroidal galaxy, Canes
Venatici, at a distance of $\sim 200$ kpc. All this suggests that
there remain unknown Milky Way companions and that \change{further
systematic surveys to find them are warranted}.  Here, we describe a
simple algorithm to carry this out in SDSS DR5, and present another
strong candidate for a Milky Way satellite, lying in the constellation
of Bo{\"o}tes. The properties of this object are somewhat unusual.

\begin{figure*}[t]
\begin{center}
\plotone{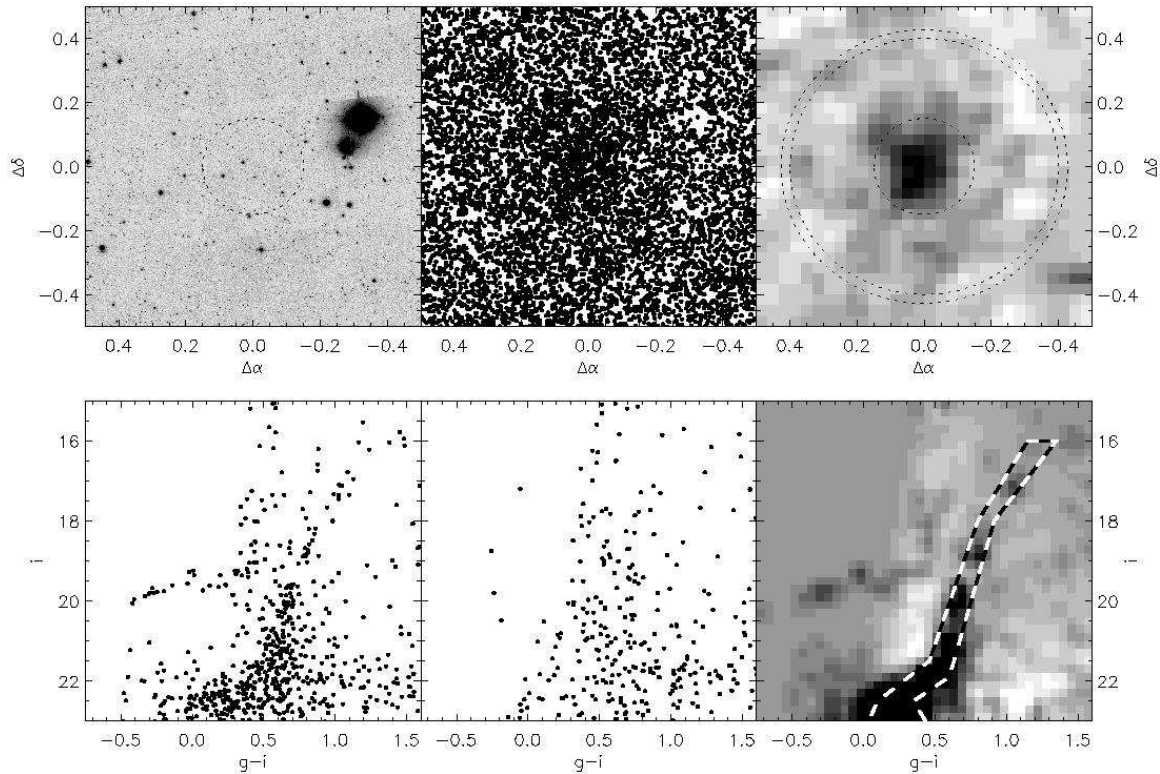}
\caption{The Bo{\"o}tes Satellite: {\it Upper left:} Combined SDSS
$g,r,i$ images of a $1^{\circ} \times 1^{\circ}$ field centered on the
overdensity. $\Delta \alpha$ and $\Delta \delta$ are the relative
offsets in right ascension and declination, measured in
arcdegrees. The dotted circle indicates a radius of
$0.15^{\circ}$. {\it Upper middle:} The spatial distribution of all
objects classified as stars in the same area. {\it Upper right:}
Binned spatial density of all stellar objects. The inner dotted circle
marks a radius of $0.15^{\circ}$ and encloses the same area as the two
outer circles, which have a radius of $0.4^{\circ}$ and $0.43^{\circ}$
respectively. Bins are $0.033^\circ \times 0.033^\circ$, smoothed with
a Gaussian with a FWHM of $0.1^\circ$.  {\it Lower left:} CMD of all
stellar objects within the inner $0.15^\circ$ radius circle. There is
a clear red giant and horizontal branch, even without removal of field
contamination. {\it Lower middle:} Control CMD, showing all stellar
objects in the annulus between $0.4^\circ$ and $0.45^{\circ}$ of the
center. {\it Lower right:} A color-magnitude density plot (Hess
diagram), showing the inner CMD minus the control CMD, normalized to
the number of stars in each CMD. A mask is shown around the
satellite's sequence.\label{fig:boo_disc}}
\end{center}
\end{figure*}

\begin{deluxetable}{lc}
\tablecaption{Properties of the Bo\"{o}tes Satellite \label{tbl:pars}}
\tablewidth{0pt} \tablehead{ \colhead{Parameter\tablenotemark{a}} &
{~~~ } } \startdata Coordinates (J2000) & \coords \\
Coordinates (Galactic) & $\ell = 358.1^\circ$, $b= 69.6^\circ$ \\
 Position Angle & $10^{\circ} \pm 10^{\circ}$\\
 Ellipticity & $0.33$\\
 $r_h$ (Plummer) & $13\farcm0 \pm 0\farcm7$\\
$r_h$ (Exponential) & $12\farcm6 \pm 0\farcm7$\\
 A$_{\rm V}$ & $0\fm06$ \\
 $\mu_{\rm 0,V}$ (Plummer) & $28\fm3 \pm 0\fm5$\\
 $\mu_{\rm 0,V}$ (Exponential) & $27\fm8 \pm 0\fm5$\\
V$_{\rm tot}$ & $13\fm6 \pm 0\fm5$\\
(m$-$M)$_0$ & $18\fm9 \pm 0\fm20$\\
 M$_{\rm tot,V}$ & $-5\fm8 \pm 0\fm5$ \enddata
\tablenotetext{a}{Surface brightnesses and integrated magnitudes are
corrected for the mean Galactic foreground reddenings, A$_{\rm V}$,
shown.}
\label{tab:struct}
\end{deluxetable}
\section{Data and Discovery}

SDSS imaging data are produced in five photometric bands, namely $u$,
$g$, $r$, $i$, and $z$~\citep{Fu96,Gu98,Gu06,Ho01}.  Thanks to
the efforts of many people, the data are automatically processed
through pipelines to measure photometric and astrometric properties
\citep{Lu99,St02,Sm02,Pi03,Iv04}. 

To carry out a systematic survey, the stars with $16 \le r \le 22$ are
first binned into $10'\times 10'$ regions in right ascension and
declination.  Then, a running window of size $1^\circ \times 1^\circ$
is used to compute the background. All bins that are more than $3
\sigma$ away from the background are selected.  Known satellite
galaxies and globular clusters are removed using the list
of~\citet{vand00}. Visual inspection is used to discard a few obvious
contaminants, such as resolved stellar associations in background
galaxies. All the candidates are ranked according to the
signal-to-noise. The two strongest candidates that remain are the
Canes Venatici dSph~\citep{Zu06} and the object studied in this {\it
Letter}, which is \change{named Boo} after the constellation of
Bo{\"o}tes in which it lies.

The upper left panel of Figure~\ref{fig:boo_disc} shows a grayscale
SDSS image of the sky centered on Boo. There is no obvious
object. However, on plotting the density of all objects classified by
the SDSS pipeline as stars, a curiously-shaped overdensity is readily
visible (upper middle and right panels).  Plotting these stars in a
color-magnitude diagram (CMD) reveals a clear red giant branch 
and horizontal branch (lower panels). This evidence of a localized
overdensity of stars with a distinct color-magnitude diagram suggests
that this is a new satellite -- possibly a dwarf galaxy.

\begin{figure}[th]
\plotone{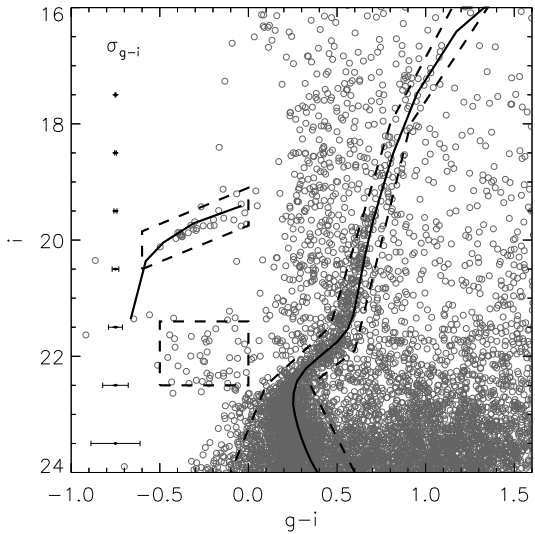}
\caption{Color-magnitude diagram of Boo derived from CTIO
data. Overplotted is the ridge-line for the old, metal-poor globular
cluster M92. The dashed lines are used to select stars belonging to
the main sequence, giant branch and horizontal branch of the
satellite. For each magnitude bin, the mean color error is shown on
the left-hand side.}
\label{fig:boo_cmd}
\end{figure}

\begin{figure}[th]
\begin{center}
\includegraphics[width=0.5\textwidth]{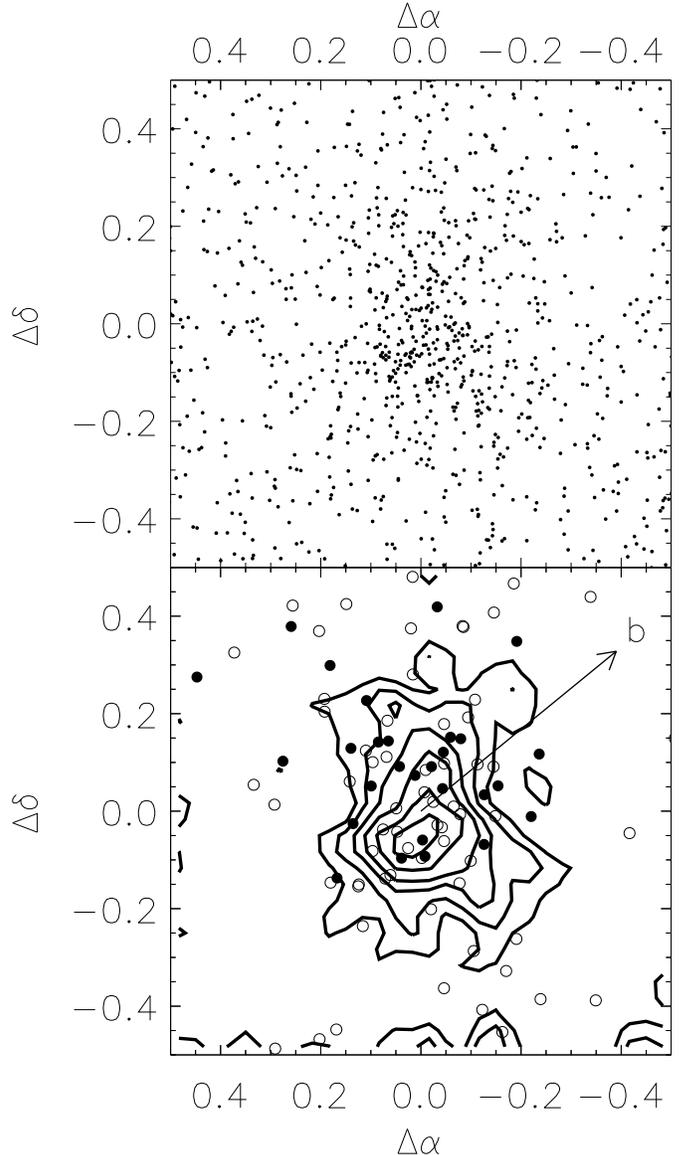}
\caption{Morphology of Boo: {\it Upper:} The spatial distribution of
SDSS stars selected from CMD regions marked with dashed lines in
Fig.~\ref{fig:boo_disc}.  {\it Lower:} A contour plot of the spatial
distribution of Boo's stars; candidate blue horizontal branch stars
and blue stragglers (from the boxes in the CMD) are overplotted with
black dots and open circles. The contours are 1.5, 3, 5, 7, 10 and 13$\sigma$
above the background level. The direction of increasing Galactic
latitude is marked by an arrow. \label{fig:boo_morph}}
\end{center}
\end{figure}

\begin{figure}[th]
\includegraphics[angle=-90,width=0.45\textwidth]{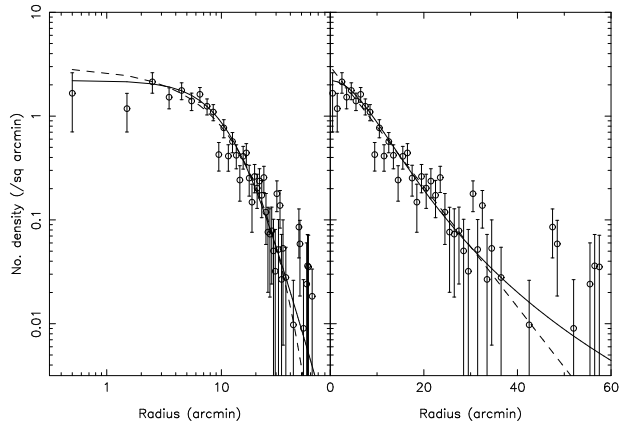}
\caption{Profile of Boo, showing the stellar density in elliptical
annuli as a function of mean radius. The left panel is logarithmic in
both axes, and the right panel is linear in radius. The overplotted
lines are fitted Plummer (solid) and exponential (dashed)
profiles.\label{fig:boo_prof}}
\end{figure}

\begin{figure}[th]
\plotone{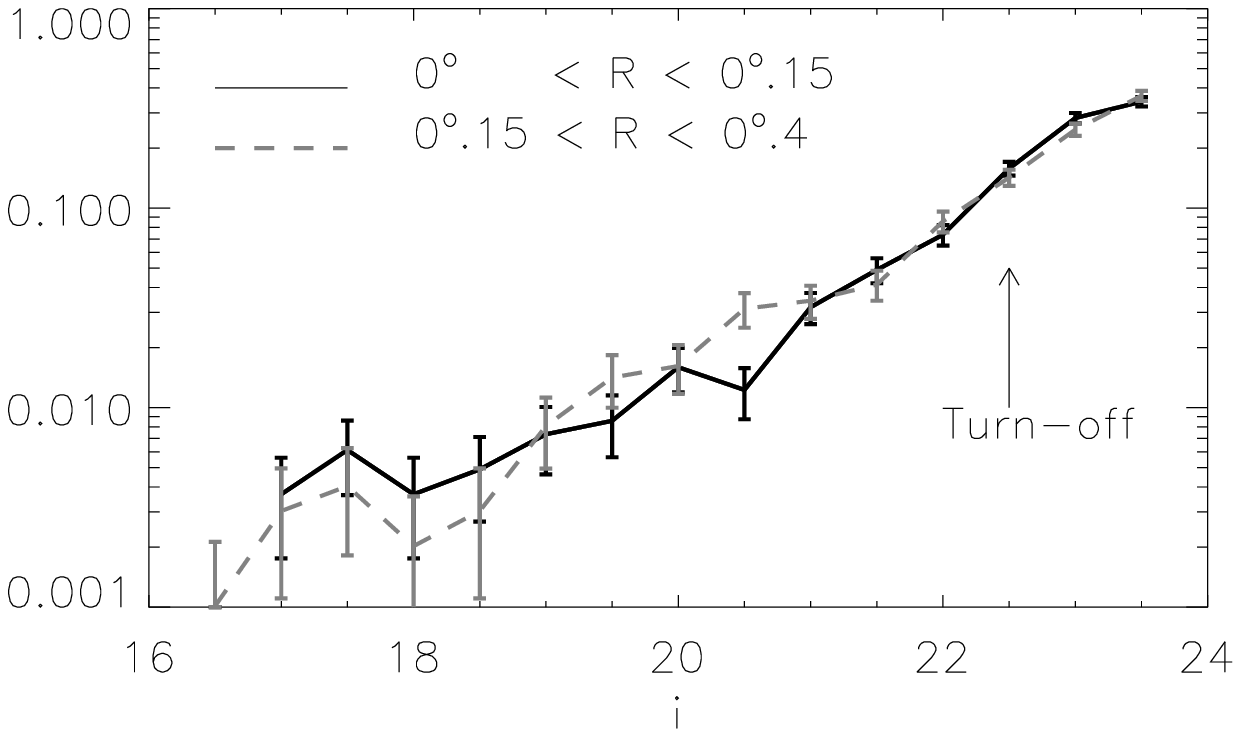}
\caption{Normalized luminosity functions of the inner (solid) and
outer (dashed) parts of Boo, constructed with the CTIO data. The 
main sequence turn-off is indicated by the arrow.
\label{fig:interrupted}}
\end{figure}

\section{Physical Properties and Stellar Population}

Follow-up observations of Boo were made on 25 Feb 2006 and 7 Mar 2006 (UT)
with the 4m Blanco telescope at Cerro Tololo Inter-American
Observatory in Chile, using the MOSAIC-II CCD camera.  This comprises 8
2k$\times$4k pixel SITe CDDs, with a field of view 36$ \times$36
arcminutes and a scale of 0.27 arcseconds per pixel at the image
centre. Boo was observed in the $g$ and $i$ bands, with exposure
times of 3$\times$360s in each filter on the first night and 3$
\times$600s in each filter on the second night, for a total exposure
of 2880s in each filter. The telescope was offset ($\approx$30
arcsecond) between exposures.

Data were processed in Cambridge using a general purpose pipeline for
processing wide-field optical CCD data (Irwin \& Lewis 2001). Images
were de-biased and trimmed, and then flatfielded and gain-corrected to
a common internal system using clipped median stacks of nightly
twilight flats.  The $i-$band images, which suffer from an additive
fringing component, were also corrected using a fringe frame computed
from a series of long $i-$band exposures taken during the night.

For each image frame, an object catalogue was generated using the
object detection and parameterisation procedure discussed in Irwin et
al. (2004).  Astrometric calibration of the invididual frames is based
on a simple Zenithal polynomical model derived from linear fits
between catalogue pixel-based coordinates and standard astrometric
stars derived from on-line APM plate catalogues.  The astrometric
solution was used to register the frames prior to creating a deep
stacked image in each passband.  Object catalogues were created
from these stellar images and objects morphologically classified as
stellar or non-stellar (or noise-like).  The detected objects in each
passband were merged by positional coincidence (within $1''$) to
form a $g$,$i$ combined catalogue. This catalogue was photometrically
calibrated onto the SDSS system using the overlap with the SDSS
catalogues.

Figure~\ref{fig:boo_cmd} shows a CMD constructed from the CTIO
data. The fiducial ridgeline of the metal-poor globular cluster M92
([Fe/H] $\sim -2.3$) from \citet{Clem05} is overplotted on the
CMD. The horizontal branch is well-matched, though Boo's giant branch
and MSTO are slightly blueward of the M92 isochrone. This is
consistent with Boo being somewhat younger and slightly more
metal-poor than M92.  Note in particular the narrowness of the giant
branch of the CMD in Figure~\ref{fig:boo_cmd}, as evidenced by the
mean color error bar shown on the left-hand side.  This is
characteristic of single epoch stellar populations, which are normally
associated with globular clusters. \change{However, although most dSph
galaxies show evidence of multiple stellar populations, a few are
known whose CMDs possess narrow red giant branches, such as Ursa Minor
and Carina~\citep{Be00}.  In both cases, the narrow red giant branch
is nonetheless consistent with a number of epochs of star
formation~\citep[see e.g.,][]{Sh01,Ko06}.} From the fact that the
horizontal branch of M92 provides a good fit, we can estimate the
distance of Boo. The distance modulus is $(m-M)_0 \sim 18.9 \pm 0.2$,
corresponding to $\sim 60 \pm 6$ kpc, where the error bar includes the
uncertainty based on differences in stellar populations of Boo and M92,
as well as the uncertainty in the distance of M92 . Note also that
there is a prominent clump in the CMD below the horizontal branch,
where Boo's blue straggler population resides.

Even though the CMD resembles that of a globular cluster, this is
emphatically not the case for the object's morphology and size.  To
select candidate stars, we use the boundaries marked by the dashed
lines on the CMD, which wrap around the satellite sequence.
\change{SDSS data are used here as the CTIO data are largely confined
to the inner parts of Boo.} The locations of SDSS stars with $r < 23$
lying within the boundaries are plotted in the top panel of
Fig.~\ref{fig:boo_morph}.  These objects are binned into $30 \times
30$ bins, each $0.033^\circ \times 0.033^\circ$, and smoothed with a
Gaussian with FWHM of $0.067^\circ$ to yield the plot in the lower
panel. The density contours, representing 1.5, 3, 5, 7, 10 and 13$\sigma$
above the background level, \change{are elongated and irregular} --
more so than even the most irregular of the Galactic dSphs, Ursa
Minor~\citep[see e.g.,][]{Pa03}.  The black dots are candidate blue
horizontal branch stars and open circles are blue stragglers. The
spatial distribution of both populations is roughly consistent with
the underlying density contours and shows the same tail-like
extensions. There are hints that Boo could be a much larger object, as
the blue horizontal branch and straggler population extend beyond the
outermost contours.

To estimate the properties listed in Table~\ref{tab:struct}, we use
the SDSS data shown in Figure~\ref{fig:boo_morph} to derive the
centroid from the density-weighted first moment of the distribution,
and the average ellipticity and position angle using the three
density-weighted second moments \citep[e.g.,][]{St80}. The radial
profile shown in Figure~\ref{fig:boo_prof} is derived by computing the
average density within elliptical annuli after first subtracting a
constant asymptotic background level (0.2 arcminute$^{-2}$) reached at
large radii.  We then fit the radial profile with standard Plummer and
exponential laws ~\citep[Figure~\ref{fig:boo_prof}, see also][]{Ir95}.
The best-fitting position angle, ellipticity and half-light radius are
listed in Table~\ref{tab:struct}. At a distance of $\sim 60$ kpc, the
half-light radius of $13\farcm0$ corresponds to $\sim 220$ pc. This is
the typical scale length of the Galactic dSph galaxies, and a factor
of $\sim 10$ times larger than the scale length of the largest
Galactic globular clusters.  Note that neither the Plummer nor the
exponential laws provide \change{exceptional} fits to the data -- in
particular, the center of the object is not well-fitted and appears to
lack a clearly defined core. Although Boo appears superficially
somewhat similar to Willman 1, it is substantially larger and more
luminous. Willman 1 has a characteristic scale length of only $\sim
20$ pc and an absolute magnitude of M$_{\rm tot,V} \sim -2\fm5$.
\change{The stellar populations are also different -- for example,
Willman 1 has no red giant or horizontal branch stars~\citep{Wi06}.}

The overall luminosity is computed by masking the stellar locus of Boo
in the CMD in Figure 1 and computing the total flux within the mask
and within the elliptical half-light radius.  A similar mask, but
covering a larger area to minimize shot-noise, well outside the main
body of Boo is scaled by relative area and used to compute the
foreground contamination within the half-light radius.  After
correcting for this contamination, the remaining flux is scaled to the
total, assuming the fitted profiles are a fair representation of the
overall flux distribution. We also apply a correction of $0.3$
magnitudes for unresolved/faint stars, based on the stellar luminosity
functions of other low metallicity, low surface brightness dSphs. The
resulting luminosity estimate is M$_{\rm tot,V} \sim -5\fm8$. Applying
our procedure to the Ursa Major dSph, discovered by~\citet{Wi05a},
gives M$_{\rm tot,V} \sim -5\fm5$. We conclude that Boo is, within the
uncertainties, comparable in faintness to Ursa Major.

We argue that Boo is not a tidally disrupted globular cluster as
follows.  First, it is much too extended. If a globular cluster is
tidally disrupted, its half-light radius may increase somewhat, but it
does not grow to such an immense radius as $\sim 220$ pc. Second, for
it to be a destroyed globular cluster, Boo would have to be on a
plunging radial orbit. Then the outermost isodensity contours, which
should be aligned with the direction of the proper motion, should point
towards the Galactic center. This is not the case, as judged from
Fig.~\ref{fig:boo_morph}.  Third, globular clusters often show
evidence for mass segregation driven by internal dynamical
evolution~\citep[see e.g.,][]{Ko04}. Accordingly,
Figure~\ref{fig:interrupted} shows normalized luminosity functions
(LFs) for the inner and outer parts of Boo constructed with the CTIO
data. There is no evidence for substantial mass segregation in Boo
with the present data. However, this data are largely restricted to
stars of similar mass, and deeper data are required to give a
conclusive result.

\section{Conclusions}

We have discovered a new companion to the Milky Way galaxy in the
constellation of Bo{\"o}tes.  The object has a globular cluster-like
CMD, dominated by an old, metal-poor stellar population. With a
characteristic scale length of 220 pc, the size of the object is
typical of the Galactic dwarf spheroidal satellites. The irregular
nature of the density contours suggests that it may be undergoing
tidal disruption. If, as seems likely, it is a dwarf galaxy, then Boo
is one of the faintest ($M_V \sim -5\fm8$) so far discovered.

\acknowledgments
We thank James Clem for providing the data on M92 used in the paper.
Funding for the SDSS and SDSS-II has been provided by the Alfred P.
Sloan Foundation, the Participating Institutions, the National Science
Foundation, the U.S. Department of Energy, the National Aeronautics and
Space Administration, the Japanese Monbukagakusho, the Max Planck
Society, and the Higher Education Funding Council for England. The SDSS
Web Site is http://www.sdss.org/.                                                                               
The SDSS is managed by the Astrophysical Research Consortium for the
Participating Institutions. The Participating Institutions are the
American Museum of Natural History, Astrophysical Institute Potsdam,
University of Basel, Cambridge University, Case Western Reserve
University, University of Chicago, Drexel University, Fermilab, the
Institute for Advanced Study, the Japan Participation Group, Johns
Hopkins University, the Joint Institute for Nuclear Astrophysics, the
Kavli Institute for Particle Astrophysics and Cosmology, the Korean
Scientist Group, the Chinese Academy of Sciences (LAMOST), Los Alamos
National Laboratory, the Max-Planck-Institute for Astronomy (MPIA), the
Max-Planck-Institute for Astrophysics (MPA), New Mexico State
University, Ohio State University, University of Pittsburgh,
University of Portsmouth, Princeton University, the United States
Naval Observatory, and the University of Washington.

\end{document}